\documentclass[fleqn,usenatbib]{mnras}

\usepackage{newtxtext,newtxmath,url}

\usepackage[T1]{fontenc}

\DeclareRobustCommand{\VAN}[3]{#2}
\let\VANthebibliography\thebibliography
\def\thebibliography{\DeclareRobustCommand{\VAN}[3]{##3}\VANthebibliography}

\usepackage{graphicx}	
\usepackage{amsmath}	

\usepackage{hyperref}
\usepackage{rotating}
\usepackage{longtable}
\usepackage{multirow}
\usepackage[normalem]{ulem}
\usepackage[flushleft]{threeparttable}
\usepackage{bm}
\usepackage{makecell}
\usepackage{array}

\usepackage{gensymb}
\usepackage{ragged2e}

\usepackage{xcolor}

\title[The joint detection rate of sGRBs and GWs]{Constraining the Luminosity Function and Delay-Time Distribution of Short Gamma-Ray Bursts for Multimessenger Gravitational-Wave Detection Rate Estimation}

\author[C.Y. Gao et al.]{
Chong-Yu Gao,$^{1,2}$
Jun-Jie Wei$^{1,2}$\thanks{E-mail: jjwei@pmo.ac.cn}
and Hou-Dun Zeng$^{1,3}$
\\
$^{1}$Purple Mountain Observatory, Chinese Academy of Sciences, Nanjing 210023, China\\
$^{2}$School of Astronomy and Space Sciences, University of Science and Technology of China, Hefei 230026, China\\
$^{3}$Key Laboratory of Dark Matter and Space Astronomy Purple Mountain Observatory, Chinese Academy of Sciences Nanjing, 210023, China
}

\date{Accepted XXX. Received YYY; in original form ZZZ}

\pubyear{\the\year{}}

\begin{document}

\label{firstpage}
\pagerange{\pageref{firstpage}--\pageref{lastpage}}
\maketitle

\begin{abstract}
In this work, we analyze the most recent short gamma-ray burst (sGRB) sample detected by the \emph{Fermi} satellite 
to reassess the sGRB luminosity function and formation rate. Using the empirical redshift-luminosity 
correlation, we first determine the pseudo redshifts of 478 sGRBs. Then, we use 
the maximum likelihood method to constrain the luminosity function and formation rate of sGRBs under various 
delay-time distribution models, finding the Gaussian delay model statistically preferred over the power-law and 
lognormal delay models based on the Deviance Information Criterion. The local formation rate of sGRBs is
$R_{\mathrm{sGRB}}(0)=1.37_{-0.27}^{+0.30}$ $\mathrm{Gpc^{-3}\,yr^{-1}}$, largely independent of the adopted delay-time 
distribution model. Additionally, we investigate the potential for joint detection of sGRBs and their gravitational wave
(GW) counterparts from binary neutron star mergers using both current and future GRB and GW facilities. For sGRB detection, 
we consider three existing satellites: \emph{Fermi}, the Space-based multi-band astronomical Variable Objects Monitor
(\emph{SVOM}), and the Einstein Probe (\emph{EP}). For GW detection, we examine two International GW Networks (IGWN):
a four-detector network consisting of LIGO Hanford, Livingston, Virgo, and KAGRA (IGWN4) and an upcoming five-detector
network that includes these four detectors plus LIGO India (IGWN5). Incorporating the angular dependence of
sGRB jet emission energy, our results show that for different delay-time
distribution models, the joint sGRB and GW detection rates for \emph{Fermi}, \emph{SVOM}, and \emph{EP} with IGWN4 (IGWN5)
lie within 0.19--0.27 $\mathrm{yr^{-1}}$ (0.93--1.35 $\mathrm{yr^{-1}}$), 0.07--0.10 $\mathrm{yr^{-1}}$ (0.51--0.79
$\mathrm{yr^{-1}}$), and 0.01--0.03 $\mathrm{yr^{-1}}$ (0.15--0.27 $\mathrm{yr^{-1}}$), respectively.
\end{abstract}




\section{Introduction}
Gamma-ray bursts (GRBs) are the most energetic explosions in the Universe. They are typically divided into two groups 
based on a critical duration value of $T_{90}=2$ seconds \citep{K1993}. Long GRBs are generally believed to originate 
from the core collapses of massive stars \citep{W1993}, whereas short GRBs (sGRBs) are thought to result from the mergers 
of two compact objects, such as neutron stars (NSs) or black holes (BHs) \citep{1989Natur.340..126E}.
The multimessenger observations of the gravitational wave event (GW170817) and its associated sGRB (GRB 170817A)
from a binary neutron star (BNS) merger have provided compelling evidence supporting the compact star merger origin of 
sGRBs \citep{ab2017,Go2017,Sa2017}. Moreover, the abundant multiwavelength follow-up observations of the afterglow of 
GRB 170817A also spurred extensive research into the jet structure and the radiation processes of sGRBs 
\citep{T2017,lyman2018,m2018,m2018a,m2018b,t2018,g2019,b2020,b2020b,t2020}. 

The multimessenger joint observations of gravitational waves (GWs) and sGRBs provide valuable information about 
the jet launching time and the propagation of the jet through the ejecta. Therefore, these observations are 
an ideal probe to understand not only the sGRB sources but also the jet dynamics and the emission mechanism 
\citep{2007PhR...442..166N,2014ARA&A..52...43B,2020PhR...886....1N}. However, aside from GW170817/GRB 170817A, 
no other joint GW and sGRB detections have been identified, which hinders the advancement of our understanding 
of the sGRB origins and the jet mechanism.

The formation rate of sGRBs is directly related to the expected number of GW events associated with sGRBs for the current
and upcoming high-energy GRB and GW facilities. However, the number of sGRBs with redshift measurements is very 
small, making it difficult to estimate the sGRB formation rate. Due to the small sample size, previous studies estimated the local
sGRB formation rate to range from $\mathrm{0.1\,Gpc^{-3}\,yr^{-1}}$ to 
$\mathrm{400\,Gpc^{-3}\,yr^{-1}}$ \citep{gp2005,2006A&A...453..823G,2006ApJ...650..281N,2009A&A...498..329G,2011A&A...529A..97D,
2013ApJ...767..140P,2014MNRAS.437..649S,2015ApJ...812...33S,gir2016,liu2019,2021ApJ...914L..40D,2023ApJ...949...15D}.
Using the empirical correlation between the spectral peak energy $E_{p}$ and the peak luminosity $L$
\citep{yo2004,2013MNRAS.431.1398T}, \cite{Yonetoku2014} determined the pseudo redshifts of those sGRBs without a known
redshift, thereby enlarging the available sGRB sample. Then they derived the luminosity function (LF) and formation rate
of sGRBs by using the non-parametric Lynden-Bell $c^{-}$ method (see also \citealt{zhang2018}).
Keep in mind that the Lynden-Bell $c^{-}$ method necessitates a certain uniformity in data coverage to be applicable
\citep{1971MNRAS.155...95L}. In reality, however, sGRB data suffer from the truncation of the flux limit 
of the detector, and the detector's sensitivity is hard to parametrize, making it challenging to construct a uniformly 
distributed sample of sGRBs. Maximum-likelihood algorithms \citep{1983ApJ...269...35M} are preferable for addressing 
the sGRB problem because they are less constrained by the conditions of a given sample and offer greater 
flexibility in modeling systematic uncertainties. In this study, we follow the treatment of \cite{Yonetoku2014} and \cite{zhang2018} 
to first obtain the pseudo redshifts of sGRBs observed by \emph{Fermi} using the $E_{p}$-$L$ correlation. 
To account for systematics and unknowns in our statistical analysis, we then employ the maximum likelihood method 
to examine the enlarged sGRB sample, thereby constraining the LF and formation rate of sGRBs.

With the derived LF and formation rate of sGRBs, we can further investigate the joint detection prospects of sGRBs and
their GW counterparts using GRB and GW facilities. The success of joint GW and sGRB detections depends on the sensitivities 
and capabilities of both GW detectors and GRB missions. Currently, the fourth observation (O4) run of the International 
Gravitational Wave Network-4 (IGWN4), which includes four detectors---Advanced LIGO Livingston and Hanford
\citep{2015CQGra..32g4001L}, Advanced Virgo \citep{2015CQGra..32b4001A}, and KAGRA at design sensitivity 
\citep{2013PhRvD..88d3007A}---is ongoing. In the future, the LIGO detectors at Livingston and Hanford will be 
upgraded to A$+$ sensitivity \citep{Barsotti18}, and the planned GW detector LIGO-India \citep{Iyer11,Saleem22} 
will joint the detector network, forming the International Gravitational Wave Network-5 (IGWN5). This IGWN5 network 
will significantly enhance the GW detection capabilities of its detectors. As the sensitivity and detection horizon 
of GW detectors continue to improve significantly, the primary limitation for joint detections is likely to soon be 
the sensitivity of GRB detectors such as \emph{Fermi}. Recently, the Space-based multi-band astronomical Variable 
Objects Monitor (\emph{SVOM}; \citealt{svommission}) and the Einstein Probe (\emph{EP}; \citealt{Yuan2022fpj}) 
have been launched into space. Both satellites share the important scientific objective of detecting 
the electromagnetic counterparts of GWs. Estimating the expected detection rates of sGRB associated with GWs 
for these two missions will be crucial for subsequent observational plans and related research.

Several studies have aimed to estimate joint GW-sGRB detection rates.
\cite{2019MNRAS.485.1435H} estimated that Advanced LIGO at design sensitivity could detect up to $\sim 4\,\mathrm{yr^{-1}}$ 
joint GW-sGRB events, increasing to $\sim 10\,\mathrm{yr^{-1}}$ with the A$+$ upgrade. For the fourth observing run (O4) 
of the LIGO-Virgo-KAGRA network, \cite{2022ApJ...937...79C} projected joint detection rates of $\sim0.03\,\mathrm{yr^{-1}}$ 
and $\sim0.17\,\mathrm{yr^{-1}}$ using \emph{Swift} and \emph{Fermi}, respectively. Similarly, \cite{p2022} reported 
rates up to $\sim6\,\mathrm{yr^{-1}}$ with \emph{Fermi} during O4, while \cite{2023A&A...680A..45S} predicted 
0.2--1.3 detectable GW-sGRB events per year for \emph{Fermi}/GBM and the Advanced GW detector network during O4.
\cite{b2024} found that for different jet structure models, joint GW-sGRB detection rates for \emph{Fermi} and \emph{Swift} 
with IGWN4 (IGWN5) range from 0.07--0.62 $\mathrm{yr^{-1}}$ (0.8--4.0 $\mathrm{yr^{-1}}$) and 0.02--0.14 $\mathrm{yr^{-1}}$ 
(0.15--1.0 $\mathrm{yr^{-1}}$), respectively.

In this work, we analyze the most recent sGRB sample detected by \emph{Fermi}'s Gamma-ray Burst Monitor (\emph{Fermi}/GBM)
to determine the LF and formation rate of sGRBs through the application of the maximum likelihood method.
Then, we explore the potential for joint sGRB and GW detections from BNS mergers using existing and upcoming GRB and
GW facilities. For sGRB detection, we consider three existing satellites: \emph{Fermi}, \emph{SVOM}, and \emph{EP}.
For GW detection, we consider two GW detector networks: the current IGWN4 and the upcoming IGWN5.
Discussions have been held regarding sGRB-like emissions from NS-BH mergers; however, the theoretical aspects 
remain unclear, and there is currently no observational confirmation. Therefore, we exclude NS-BH mergers from our analysis.

This paper is organized as follows. In Section \ref{sec:samples}, we describe the \emph{Fermi}/GBM sample available to us,
and obtain the pseudo redshifts of sGRBs. In Section \ref{sec:sGRB}, the LF and formation rate of sGRBs are determined
through the maximum likelihood method. The synthetic BNS population model, the sGRB detection rates, and the prospects
of joint sGRB and GW detections are presented in Section \ref{sec:joint}. Finally, we provide a brief summary and discussion 
in Section \ref{sec:con}. Throughout this paper, we assume a standard $\Lambda$CDM cosmological model and adopt 
the parameters of $H_0=67.4$ $\mathrm{km}$ $\mathrm{s^{-1}}$ $\mathrm{Mpc^{-1}}$, $\mathrm{\Omega_{m}=0.315}$, 
and $\mathrm{\Omega _{\Lambda}=1-\mathrm{\Omega_{m}=0.315}}$ \citep{planck2020}.

\section{\emph{Fermi} GRB Sample and Redshift Estimation}
\label{sec:samples}
\subsection{Sample selection}
\begin{figure}
\vskip-0.2in
\includegraphics[width=\columnwidth]{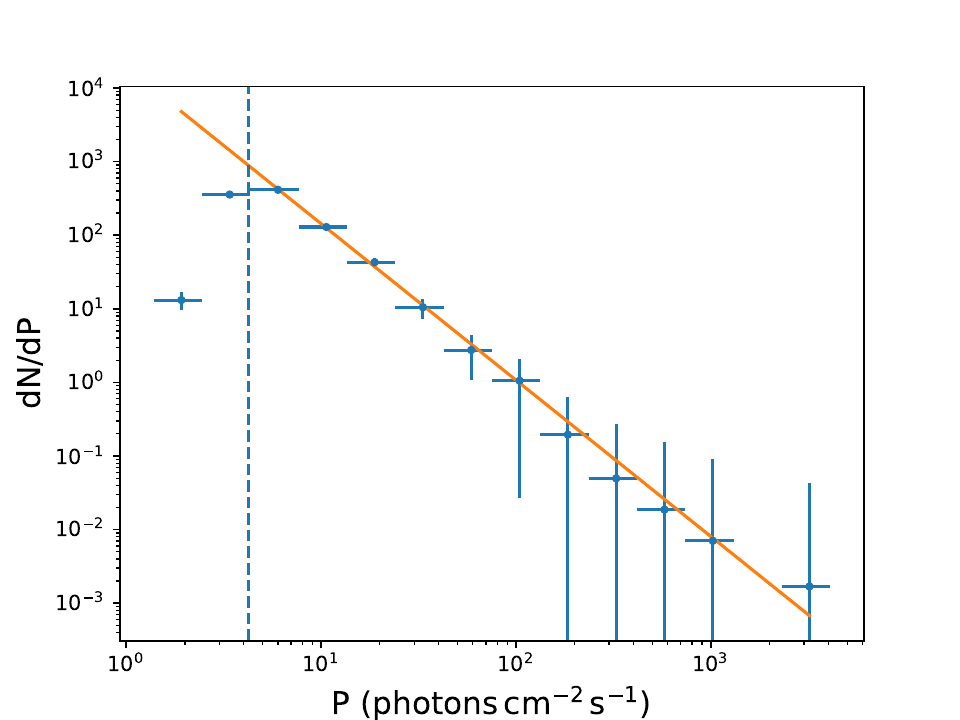}
\vskip-0.05in
    \caption{The peak-flux distribution of 553 sGRBs detected by \emph{Fermi}/GBM.
    The vertical dashed line stands for the limiting flux ($\sim4.1\,\mathrm{photons\,cm^{-2}\,s^{-1}}$) 
    above which the instrumental selection effects on incomplete sampling of faint bursts
    become negligible. The solid line represents the best power-law fit to the peak-flux distribution for $P\geq4.1$ $\mathrm{photons\,cm^{-2}\,s^{-1}}$.}
    \label{fig:Sdis}
\end{figure}

Since the launch of the \emph{Fermi} satellite on 2008 June, 3671 GRBs have been detected up to 2023 December 15.
We select 553 sGRBs with durations $T_{90}<2$ s in the observer frame. For each sGRB, we download the 64 ms 
peak photon flux $P$ (in units of $\mathrm{photons\,cm^{-2}\,s^{-1}}$) in the 10--1000 keV energy band and 
the spectral parameters (including the low- and high-energy photon indices $\alpha$ and $\beta$, 
and the observed peak energy $E_p$) from the online \emph{Fermi}/GBM burst catalog 
\citep{2014ApJS..211...12G,2014ApJS..211...13V,2016ApJS..223...28N,2020ApJ...893...46V}.\footnote{
\url{https://heasarc.gsfc.nasa.gov/W3Browse/fermi/fermigbrst.html}} Figure~\ref{fig:Sdis} shows the peak-flux
distribution of all the 553 sGRBs, ranging from 1.91 to 3198.53 $\mathrm{photons\,cm^{-2}\,s^{-1}}$.

The \emph{Fermi}/GBM trigger mechanism is complex, and its sensitivity to GRBs is difficult to 
parametrize precisely. In practice, not all faint GRBs with peak fluxes slightly above the \emph{Fermi}/GBM 
detection threshold are successfully triggered. As shown in Figure~\ref{fig:Sdis}, the peak-flux distribution 
deviates notably from an ideal power law at the lower end due to incomplete sampling of faint bursts. 
To address this, we fit the peak-flux distribution with a simple power law and find that bursts with peak fluxes 
greater than 4.1 $\mathrm{photons\,cm^{-2}\,s^{-1}}$ are well described by a power-law distribution. To avoid complications
arising from a detailed treatment of the \emph{Fermi}/GBM trigger, we focus on a sample of 512 sGRBs with peak fluxes of
$P\geq4.1$ $\mathrm{photons\,cm^{-2}\,s^{-1}}$. Indeed, above the flux limit ($P_{\rm lim}=4.1$
$\mathrm{photons\,cm^{-2}\,s^{-1}}$), the instrumental selection effects leading to incomplete sampling are negligible 
(see Figure~\ref{fig:Sdis}). Thus, we assume that the trigger efficiency for bursts with $P\geq P_{\rm lim}$
is 100\%.

\subsection{Redshift estimation for sGRBs}
\label{subsec:redshiftcal}
\begin{figure}
\vskip-0.2in
\includegraphics[width=\columnwidth]{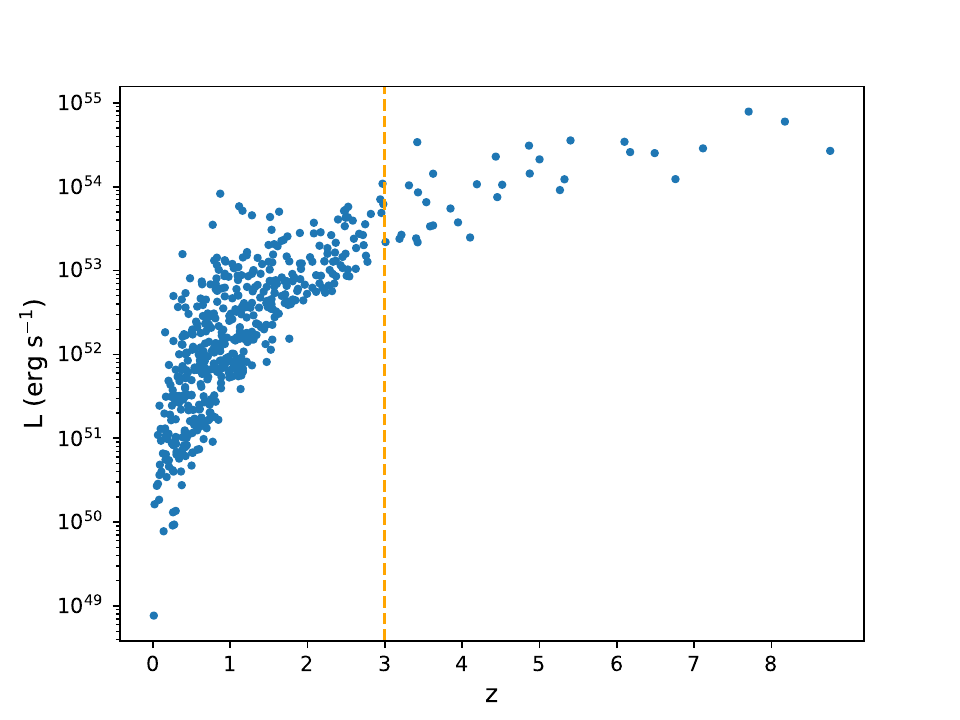}
\vskip-0.1in
    \caption{Pseudo redshift--luminosity distribution of 512 \emph{Fermi} sGRBs estimated by the $E_{p}$-$L$
    correlation. The vertical dashed line represents the truncated redshift of $z=3$, below which 478 bursts are
    used for our analysis.}
    \label{fig:L-Z}
\end{figure}

The current number of sGRBs with redshift measurements is very small. In order to study their LF
and formation rate, we determine the pseudo redshifts of sGRBs observed by \emph{Fermi}/GBM using the empirical 
correlation between the observer-frame spectral peak energy ($E_p$) and the source-frame 
peak bolometric luminosity ($L$)\citep{2013MNRAS.431.1398T,zhang2018}. This correlation is given by \citep{zhang2018} 
\begin{equation} 
    L=(7.15\pm0.49)\times10^{50}\biggl[\frac{E_{p}(1+z)}{100\;\mathrm{keV}}\biggr]^{1.63\pm0.03}\,,
    \label{eq:yonetoku}
\end{equation}
which can be rewritten as
\begin{equation}
    \label{eq:yonetoku2}
    \frac{d_{l}^2(z)}{(1+z)^{1.63}}=\frac{7.15\times10^{50}}{4\pi F_p}\biggl(\frac{E_p}{100\,\mathrm{keV}}\biggr)^{1.63},
\end{equation}
where $d_l$ is the luminosity distance and $F_p$ is the source-frame peak bolometric flux at 64 ms time intervals
(in units of $\mathrm{erg\,cm^{-2}\,s^{-1}}$). The pseudo redshift of each burst can be estimated using 
the spectral parameters of the Band function \citep{Band93} and the 64 ms peak photon flux $P$ in the 10--1000 keV 
observer-frame energy band. With the pseudo redshift, 
we can further calculate the peak bolometric luminosity $L$ in the 1--$10^5$ keV source-frame energy range. 
In our calculations, we adopt the spectral parameters from the \emph{Fermi}/GBM burst catalog. 
Our sample contains 181 sGRBs without spectral parameter measurements. For these events, 
we set $\alpha=-1$ and $\beta=-2.25$ \citep{2000ApJS..126...19P,2006ApJS..166..298K}, while $E_p$ is 
randomly drawn from the observed distribution of \emph{Fermi} sGRBs with spectral measurements.
\cite{2018MNRAS.473.3385P} demonstrated that this method statistically reproduces the pseudo redshifts 
of long GRBs with known redshifts. This approach is justified because \emph{Fermi}/GBM, as a wide-band detector, 
samples the $E_p$ space with minimal energy-dependent selection bias. Furthermore, \cite{p2018} applied
this method to estimate the pseudo redshifts of sGRBs. The pseudo redshift--luminosity distribution of 
512 sGRBs is shown in Figure~\ref{fig:L-Z}. Since the maximum redshift observed for sGRBs is 2.609, 
we exclude those bursts with pseudo-redshifts $z>3$. Hereafter, we use 478 sGRBs with $P\geq4.1$ 
$\mathrm{photons\,cm^{-2}\,s^{-1}}$ and $z\leq3$ for further analysis.

\begin{figure}
\vskip-0.2in
\includegraphics[width=\columnwidth]{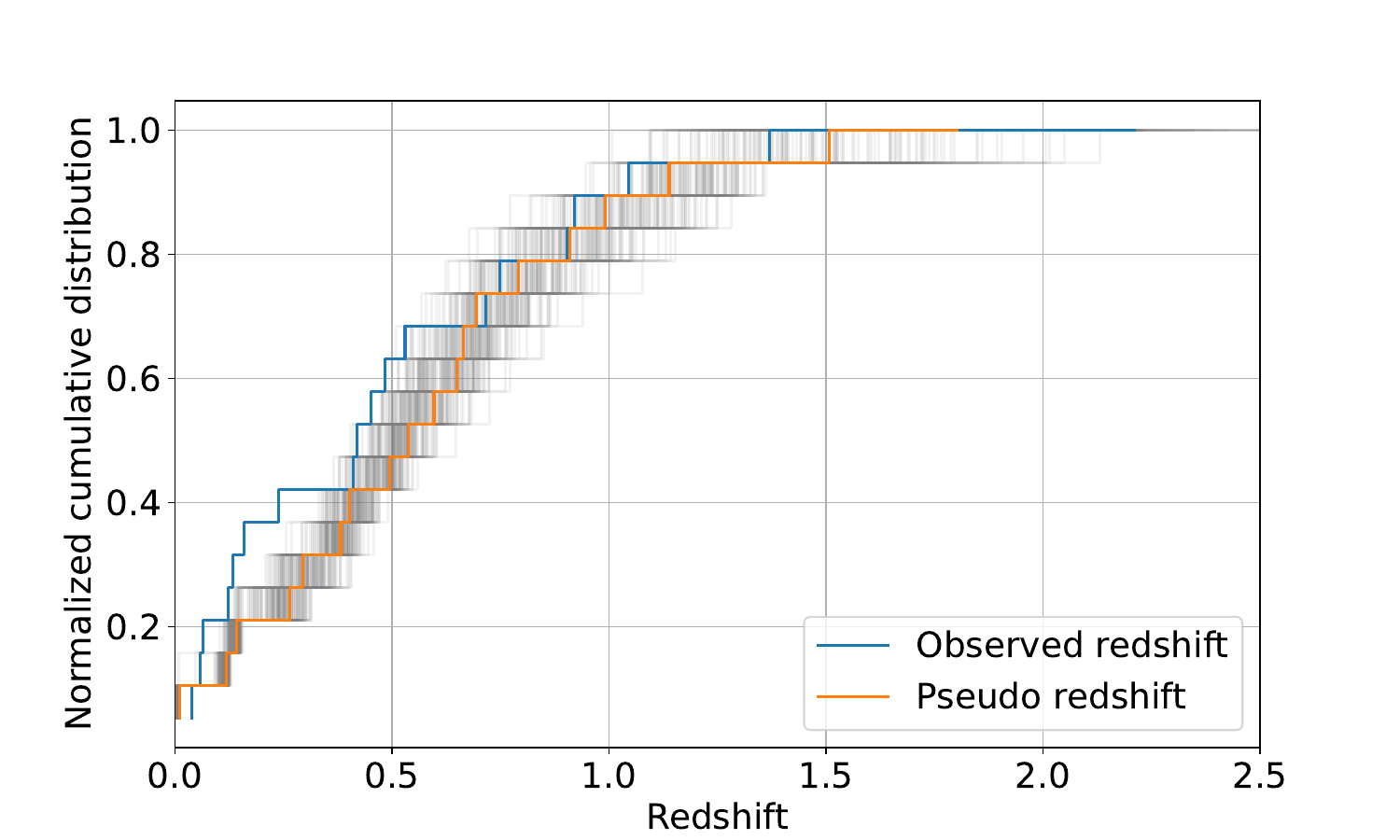}
\vskip-0.05in
    \caption{Normalized cumulative redshift distribution for 18 sGRBs with known redshifts 
    in our sample (blue line). The orange line represents the pseudo-redshift distribution for these bursts
    derived from the $E_{p}-L$ correlation. The gray lines show uncertainty regions estimated from 100 Monte Carlo simulations.}
    \label{fig:kstest}
\end{figure}

Among our final sample of 478 sGRBs, 18 bursts have spectroscopically confirmed redshifts. 
In Table~\ref{tab:redshift}, we list the measured redshifts and pseudo-redshift estimates 
for these 18 bursts. Note that among these 18 bursts, only GRB 210510806 lacks
a measured $E_p$. Following \cite{2018MNRAS.473.3385P,p2018}, we randomly sample this burst's $E_p$
from the observed $E_p$ distribution of \emph{Fermi} sGRBs with spectral measurements. 
To validate our pseudo-redshift estimates, we compare 
the cumulative distributions of measured (blue line) and pseudo-redshifts (orange line) 
in Figure~\ref{fig:kstest} and perform a Kolmogorov-Smirnov (K-S) test between these distributions.
The K-S test statistic, $P_{\rm KS}$, quantifies the likelihood that two datasets are drawn from 
the same distribution: a higher $P_{\rm KS}$-value indicates greater consistency between 
the distributions. By convention, $P_{\rm KS}>0.05$ is considered insufficient evidence 
to reject the null hypothesis, while of $P_{\rm KS}<10^{-4}$ provides strong evidence 
against the null hypothesis (i.e., the distributions differ).
The resulting $P_{\rm KS}$-value of 0.80 indicates that we cannot reject the null hypothesis. 
Moreover, we estimate an uncertainty region for the pseudo-redshift distribution of these 18 bursts. 
Redshift errors primarily originate from $E_p$ uncertainties, so we performed 100 Monte Carlo 
simulations per data point to derive cumulative distributions. Gray lines in Figure~\ref{fig:kstest} 
show these results, indicating the uncertainty region partially covers the observed distribution 
(blue line). Therefore, we conclude that the redshift distribution estimated from the $E_{p}-L$ correlation 
is broadly consistent with the observed distribution, supporting the use of $E_{p}-L$ correlation 
as a statistically valid distance indicator for sGRBs. Notably, \cite{2018MNRAS.473.3385P} 
demonstrated that random $E_p$ sampling statistically replicates the redshift distribution of long GRBs. 
Figure~\ref{fig:kstest} confirms this approach also works for sGRBs: their redshift distribution 
is replicated through $E_p$ uncertainty propagation and random $E_p$ draws for spectrally unmeasured bursts.

In our pseudo-redshift estimation, we make no assertion regarding the accuracy of individual values. 
We do not assess pseudo-redshift accuracy per GRB, as our analysis relies exclusively on statistical distribution 
comparisons. This approach assigns pseudo redshifts to bursts in a collective statistical context only. 
Nevertheless, this limitation does not invalidate our methodology, since we use pseudo-redshifts and 
derived luminosities solely as ensemble quantities for modeling the sGRB LF and formation rate.

\begin{table}
\renewcommand{\arraystretch}{1.6}
\caption{The measured redshifts and pseudo-redshift estimates of 18 sGRBs.}
\label{tab:redshift}
{\begin{tabular}{ccc}\hline
GRB Name & Measured Redshift$^{a}$  & Pseudo Redshift$^{b}$  \\ 
\hline                            
GRB 090927422 & 1.37 & 0.66 \\ 
GRB 100625773 & 0.45 & 0.66 \\ 
GRB 200826187 & 0.75 & 0.25 \\ 
GRB 090510016 & 0.90 & 1.14 \\ 
GRB 080905499 & 0.12 & 1.82 \\ 
GRB 231024556 & 0.06 & 0.40 \\ 
GRB 150101641 & 0.13 & 0.26 \\ 
GRB 131004904 & 0.72 & 0.79 \\ 
GRB 100117879 & 0.92 & 1.01 \\ 
GRB 100206563 & 0.41 & 0.55 \\ 
GRB 111117510 & 2.21 & 0.69 \\ 
GRB 210323918 & 0.42 & 0.91 \\ 
GRB 160624477 & 0.48 & 1.55 \\ 
GRB 100216422 & 0.04 & 0.28 \\ 
GRB 220211047 & 0.53 & 0.36 \\ 
GRB 201221963 & 1.05 & 0.91 \\ 
GRB 210510806 & 0.22 & 1.11$^{c}$ \\ 
GRB 160821937 & 0.16 & 0.59 \\ \hline
\end{tabular}}
\begin{description}
  \item[\emph{Note.}] {$^{a}$The measured redshifts are obtained from \url{https://user-web.icecube.wisc.edu/~grbweb_public/Summary_table.html}. $^{b}$Pseudo-redshift estimation involves statistical analysis, and measured versus pseudo redshifts are not meant to be equal since pseudo redshifts derive from the empirical $E_p$-$L$ correlation and, for bursts without spectral measurements, random $E_p$ sampling. $^{c}$For GRB 210510806, its reported pseudo redshift corresponds to a single random $E_p$ draw.}
\end{description}
\end{table}

\section{sGRB Luminosity Function and Formation rate}
\label{sec:sGRB}

\subsection{Analysis method}
In this work, we use the maximum likelihood method proposed by \cite{1983ApJ...269...35M} to constrain 
the sGRB LF and redshift distribution. The likelihood function is defined as (e.g.,
\citealt{1998ApJ...496..752C,2006ApJ...643...81N,2009ApJ...699..603A,2012ApJ...751..108A,2010ApJ...720..435A,
2019MNRAS.488.4607L,lan2021,2022ApJ...938..129L})
\begin{equation}
    \mathcal{L}=\exp\begin{pmatrix}-N_{\exp}\end{pmatrix}\prod_{i=1}^{N_{\mathrm{obs}}}\Phi(L_{i},\,z_{i},\,t_{i})\,,
    \label{eq:mle}
\end{equation}
where $N_{\rm exp}$ is the expected number of sGRB detections, $N_{\rm obs}$ is the observed number of the sample, 
and $\Phi(L,\,z,\,t)$ denotes the observed rate of sGRBs per unit time at redshift $z\sim z+\mathrm{d}z$ with luminosity
$L\sim L+\mathrm{d}L$, which can be written as
\begin{equation}
    \label{eq:totaldis}
    \Phi(L,\,z,\,t)=\frac{\mathrm{d}^{3}N}{\mathrm{d}t\mathrm{d}L\mathrm{d}z}=f_{\rm sky}f_{\rm not\_SAA}\frac{R_{\rm sGRB}(z)}{1+z}\frac{\mathrm{d}V(z)}{\mathrm{d}z}\phi(L)\,,
\end{equation}
where $f_{\rm sky}=0.7$ is the sky-coverage fraction of \emph{Fermi}/GBM,\footnote{\emph{Fermi} is a low earth-orbit 
satellite that suffers from limited sky coverage because the Earth blocks $\sim30\%$ of the sky.} 
$f_{\rm not\_SAA}=0.85$ denotes the fraction of \emph{Fermi}'s orbit outside the South Atlantic Anomaly (SAA),
$R_{\rm sGRB}(z)$ is the comoving rate density of sGRBs in units of $\mathrm{Gpc^{-3}\,yr^{-1}}$, 
the factor $(1+z)^{-1}$ reflects the cosmological time dilation, $\phi(L)$ is the normalized LF of sGRBs, 
and $\mathrm{d}V(z)/\mathrm{d}z=4\pi cd_{l}^2(z)/[H_{0}(1+z)^{2}\sqrt{\Omega_{\mathrm{m}}(1+z)^{3}+\Omega_{\Lambda}}]$ 
is the comoving volume element in a flat $\Lambda$CDM cosmology.

It is believed that sGRBs originate from the mergers of two compact stars, either NS-NS or NS-BH. A compact 
binary system undergoes a long inspiral phase before the final merger, resulting in a significant time delay 
with respect to star formation \citep{faber2012,burns2020NeutronStarMergers}. If we assume that the fraction 
of mass forming compact binary systems relative to the total mass forming newborn stars remains constant over 
cosmological time, then the sGRB rate density $R_{\rm sGRB}(z)$ can be represented as a convolution of the star 
formation rate density (SFRD) $\psi_\star(z)$ and the probability density function of the delay time distribution 
$P(\tau)$ \citep{p2018,liu2019,luo2022}:
\begin{equation}
\label{eq:formation}
    R_{\mathrm{sGRB}}(z)=f_{b}C\int_{\tau_{\rm min}}^{\tau_{\rm max}}\psi_{\star}[z'(\tau)]P(\tau) \mathrm{d}\tau\,,
\end{equation}
where $f_{b}=(1-\cos\theta_{j})$ represents the beaming factor for a top-hat jet with opening angle  $\theta_{j}$, 
and $C$ is the sGRB formation efficiency per unit available mass (in units of $\mathrm{M^{-1}_{\odot}}$).
In the maximum likelihood analysis, the parameters $C$ and $f_{b}$ are degenerate. That is, one can vary 
either $C$ or $f_{b}$, but not both. Therefore, we adopt a definition
\begin{equation}
\eta\equiv f_{b}C\,,
\end{equation}
where $\eta$ is the `$f_{b}$-free' formation efficiency, still in units of $\mathrm{M^{-1}_{\odot}}$.
Here, $z'(\tau)$ denotes the redshift at the formation of the compact binary system, while $z$ corresponds to 
the redshift of the sGRB event. Thus, the delay time between the formation and merger of the compact binary system 
is given by $\tau=t(z)-t(z')$, where $t(z)$ and $t(z')$ are the ages of the Universe at redshifts $z$ and $z'$, respectively.
In view of the fact that the progenitors of NSs usually have initial masses of 10--29 $\mathrm{M_{\odot}}$, 
and their lifetimes are 10--30 $\mathrm{Myr}$, we set the minimum delay time to be $\tau_{\rm min}=10\,\mathrm{Myr}$, 
as done by \cite{luo2022} in their treatment. The age of the Universe at redshift $z$, $t(z)$, is set as 
the maximum delay time $\tau_{\rm max}$. The SFRD $\psi_\star(z)$ can be expressed approximately by the analytical form 
\citep{yuksel2008RevealingHighredshiftStar}:
\begin{equation}
    \label{eq:sfrd}
    \psi_\star(z)=\rho_0\left[\left(1+z\right)^{3.4\omega}+\left(\frac{1+z}{5000}\right)^{-0.3\omega}+\left(\frac{1+z}{9}\right)^{-3.5\omega}\right]^{1/\omega}\;,
\end{equation}
where $\rho_{0}=0.02\,\mathrm{M_{\odot}\,Mpc^{-3}\,yr^{-1}}$ is the local SFRD, and $\omega=-10$. 

For the delay time distribution $P(\tau)$, we consider three most widely discussed models: Gaussian, Lognormal, 
and Power law, whose functional forms can be expressed as:
\begin{enumerate}
	\item Gaussian \citep{V2011}:
	\begin{equation}
	P(\tau) \propto \frac{1}{\sqrt{2\upi}\sigma_{\rm G}}\exp\left[-\frac{\left(\tau-t_{\rm G}\right)^2}{2\sigma_{\rm G}^2}\right]\,,
	\end{equation}
        where $t_{\rm G}$ and $\sigma_{\rm G}$ represent, respectively, the mean and the standard deviation of $\tau$.
	
	\item Lognormal \citep{wp2015}:
	\begin{equation}
	P(\tau) \propto \frac{1}{\sqrt{2\pi}\tau\sigma_{\rm LN}}\exp\left[-\frac{\left(\ln\tau-\ln t_{\rm LN}\right)^2}{2\sigma_{\rm LN}^2}\right]\,,
	\end{equation}
	where $\ln t_{\rm LN}$ and $\sigma_{\rm LN}$ represent, respectively, the mean and the standard deviation of $\ln\tau$.
	
	\item Power law \citep{n2006}:
	\begin{equation}
	P(\tau) \propto \tau^{-\gamma}\;,
	\end{equation}
        where $\gamma$ is the power-law index.
\end{enumerate}

For the sGRB LF $\phi(L)$, we adopt the widely recognized expression in terms of a broken power law:
\begin{equation}
\label{eq:LF}
 \phi (L)  \propto \left\{\begin{array}{l}
{\left(\frac{{{L}}}{{{L_{b}}}}\right)^{-\nu_1}};\,\,{L} \le {L_{b}}\;,\\
{\left(\frac{{{L}}}{{{L_{b}}}}\right)^{-\nu_2}};\,\,{L} > {L_{b}}\;,
\end{array} \right.
\end{equation}
where $\nu_1$ and $\nu_2$ are the power-law indices before and after the break luminosity $L_b$. The normalization
constant of the LF is calculated between the minimum and maximum luminosities, $L_{\rm min}=10^{49}\,\mathrm{erg\,s^{-1}}$
and $L_{\rm max}=10^{55}\,\mathrm{erg\,s^{-1}}$ \citep{2018MNRAS.475.1331T}.

Finally, when considering \emph{Fermi}/GBM having a flux threshold of $P_{\rm lim}=4.1$ $\mathrm{photons\,cm^{-2}\,s^{-1}}$
in the 10--1000 keV energy band, the expected number of sGRBs should be
\begin{equation}
\begin{aligned}
N_{\rm exp}=T f_{\rm sky} f_{\rm not\_SAA}\int_{0}^{z_{\rm max}}\frac{R_{\rm sGRB}(z)}{1+z}\frac{\mathrm{d}V(z)}{\mathrm{d}z}\mathrm{d}z\\
\int_{\max[L_{\rm min},\,L_{\rm lim}(z)]}^{L_{\rm max}}\phi(L)\mathrm{d}L \,,
\end{aligned}
\label{eq:N}
\end{equation}
where $T\sim15.5\,\mathrm{yr}$ is the observational period of \emph{Fermi} that covers our sample, and $z_{\rm max}=3$
corresponds to the maximum redshift of our sample. For a given redshift $z$, the luminosity threshold can be determined by 
\begin{equation}
    \label{eq:L}
     L_{\rm lim}(z)=4\pi d_{L}^{2}(z)P_{\rm lim}k(z)\;,
\end{equation}
where $k(z)$ is the spectral $k$-correction factor converting \emph{Fermi}/GBM's observed energy band of 10--1000 keV
(denoted by $E_{\rm min}$--$E_{\rm max}$ keV) into the source-frame band of $1$--$10^{5}$ keV. For \emph{Fermi} bursts, 
the flux threshold $P_{\rm lim}$ is given in units of $\mathrm{photons\,cm^{-2}\,s^{-1}}$, hence
\begin{equation}
    \label{eq:k1}  k(z)=\frac{\int_{1/(1+z)\,\mathrm{keV}}^{10^{5}/(1+z)\,\mathrm{keV}}EN(E)\mathrm{d}E}{\int_{E_{\rm min}}^{E_{\rm max}}N(E)\mathrm{d}E}\,,
\end{equation}
where $N(E)$ is the observed photon spectrum. While for other detectors with $P_{\rm lim}$ given in
$\mathrm{erg\,cm^{-2}\,s^{-1}}$, the $k$-correction factor should be rewritten as follows:
\begin{equation}
    \label{eq:k2}  k(z)=\frac{\int_{1/(1+z)\,\mathrm{keV}}^{10^{5}/(1+z)\,\mathrm{keV}}EN(E)\mathrm{d}E}{\int_{E_{\rm min}}^{E_{\rm max}}EN(E)\mathrm{d}E}\,.
\end{equation}
To describe $N(E)$, we adopt a typical Band spectrum, characterized by 
photon spectral indices of $\alpha=-1$ and $\beta=-2.25$ \citep{Band93,2000ApJS..126...19P,2006ApJS..166..298K}.
For a given $L$, the observer-frame spectral peak energy $E_p$ is estimated through the empirical $E_{p}-L$ correlation 
(i.e., Equation~\ref{eq:yonetoku}).

\subsection{Results}
For a given delay-time distribution model, we optimize the free parameters by maximizing the likelihood function
(Equation~\ref{eq:mle}). The delay time distribution $P(\tau)$ has one or two free parameters (e.g., $t_{\rm G}$ 
and $\sigma_{\rm G}$ for Gaussian delay, $t_{\rm LN}$ and $\sigma_{\rm LN}$ for lognormal delay, or $\gamma$ for
power-law delay). The sGRB LF $\phi(L)$ has three free parameters ($\nu_1$, $\nu_2$, and $L_b$). Additionally, 
the sGRB rate density $R_{\rm sGRB}(z)$, given by Equation~(\ref{eq:formation}), depends on the sGRB formation 
efficiency $\eta$, resulting in a total of five or six free parameters. For each model, the Python Markov chain 
Monte Carlo (MCMC) module, EMCEE \citep{Foreman2013}, is employed to derive multidimensional parameter constraints
from the observational data. 
The one-dimensional (1D) probability distributions and two-dimensional (2D) confidence regions 
(with 1-2$\sigma$ contours) for the parameters of different delay-time distribution models are shown in
Figures~\ref{fig:cornerGauss}-\ref{fig:cornerPL}.
These plots demonstrate that the sGRB formation 
efficiency ($\eta$) and the parameters of the sGRB LF ($\nu_1$, $\nu_2$, and $L_b$) are well constrained, regardless of the assumed delay-time distribution. For the power-law delay model, the power-law index ($\gamma$) is also well determined. In contrast, for the Gaussian and lognormal delay models, there is a strong degeneracy between the delay parameters: $t_{\rm G}$ and $\sigma_{\rm G}$ for the Gaussian model, and $t_{\rm LN}$ and $\sigma_{\rm LN}$ for the lognormal model. This degeneracy allows only an upper limit to be placed on $\sigma_{\rm G}$ (Gaussian) or $\sigma_{\rm LN}$ (lognormal).

Table~\ref{tab:bestfit} lists the best-fit parameters and their corresponding $1\sigma$ uncertainties 
for different models. In the last two columns, we also list the logarithmic likelihood value ($\ln \mathcal{L}$) 
and the Deviance Information Criterion (DIC) score, which can be used to statistically judge which of the models
is preferred by the observational data. For each model, the DIC score is calculated as $\text{DIC} = \overline{D(\theta)} + p_D$, where $D(\theta) = -2 \ln \mathcal{L}(\theta)$ is the deviance
of the likelihood, and $p_D = \overline{D(\theta)} - D(\overline{\theta})$. Here, \(\overline{D(\theta)}\) represents the mean deviance averaged over posterior samples, $\overline{\theta}$ denotes the posterior mean of the parameters.
Unlike the Akaike and Bayesian Information Criteria, the DIC explicitly accounts for scenarios where parameters (or combinations of parameters) are poorly constrained by observational data, a common challenge in astrophysics \citep{liddle2007}.
Given multiple different models $\mathcal{M}_1$,
$\mathcal{M}_2$, ..., $\mathcal{M}_N$, each with the corresponding DIC scores $\mathrm{
DIC_1}$, $\mathrm{DIC_2}$, ..., 
$\mathrm{DIC}_N$, the unnormalized confidence that model $\mathcal{M}_i$ is correct is the Akaike weight $\exp(-\mathrm{DIC}_i/2)$. 
Thus, the relative probability that $\mathcal{M}_i$ is statistically preferred is 
\begin{equation}
    P(\mathcal{M}_i)=\frac{\exp(-\mathrm{DIC}_i/2)}{\exp(-\mathrm{DIC}_1/2)+\cdots+\exp(-\mathrm{DIC}_N/2)}\,.
    \label{eq:aic}
\end{equation}

\begin{table*}
\centering
\renewcommand{\arraystretch}{1.4}
\caption{Best-fitting parameters for different delay-time distribution models.}
\label{tab:bestfit}
\resizebox{\textwidth}{!}{
\begin{tabular}{lccccccc}
\hline
		Delay model & Delay parameter &$\log_{10}\eta$ &  $\log_{10}L_b$  & $\nu_1$ & $\nu_2$ & $\ln \mathcal L$ & DIC\\
                  &                 &(${\rm M}_{\odot}^{-1}$)&  $(\mathrm{erg\,s^{-1}})$ &   &   &  &    \\
\hline
        Gaussian  & $t_{\rm G}=1.14_{-0.17}^{+0.02}$,\, $\sigma_{\rm G}<0.26$ & $1.45_{-0.08}^{+0.12}$ & $53.25_{-0.18}^{+0.13}$ & $1.40_{-0.04}^{+0.04}$ & $2.76_{-0.35}^{+0.64}$ & $-47486$ & 94996  \\
        Lognormal  & $t_{\rm LN}=2.14_{-0.28}^{+0.01}$,\, $\sigma_{\rm LN}<0.06$ & $1.30_{-0.06}^{+0.06}$ & $53.22_{-0.10}^{+0.16}$ & $1.37_{-0.03}^{+0.03}$ & $2.72_{-0.28}^{+0.51}$ & $-47520$ & 95053 \\
        Power law  & $\gamma = 2.04_{-0.48}^{+0.44}$ & $1.51_{-0.06}^{+0.06}$ & $53.17_{-0.09}^{+0.10}$ & $1.29_{-0.04}^{+0.04}$ & $2.63_{-0.30}^{+0.24}$ & $-47526$ & 95058\\  
  \hline
\end{tabular}
}
\begin{description}
  \item[\emph{Note.}] {$t_{\rm G}$, $t_{\rm LN}$, and $\sigma_{\rm G}$ are in units of Gyr, and $\sigma_{\rm LN}$ is in units of $\ln \mathrm{Gyr}$.}
\end{description}
\end{table*}

\begin{figure*}
\vskip-0.1in
\includegraphics[width=0.45\textwidth]{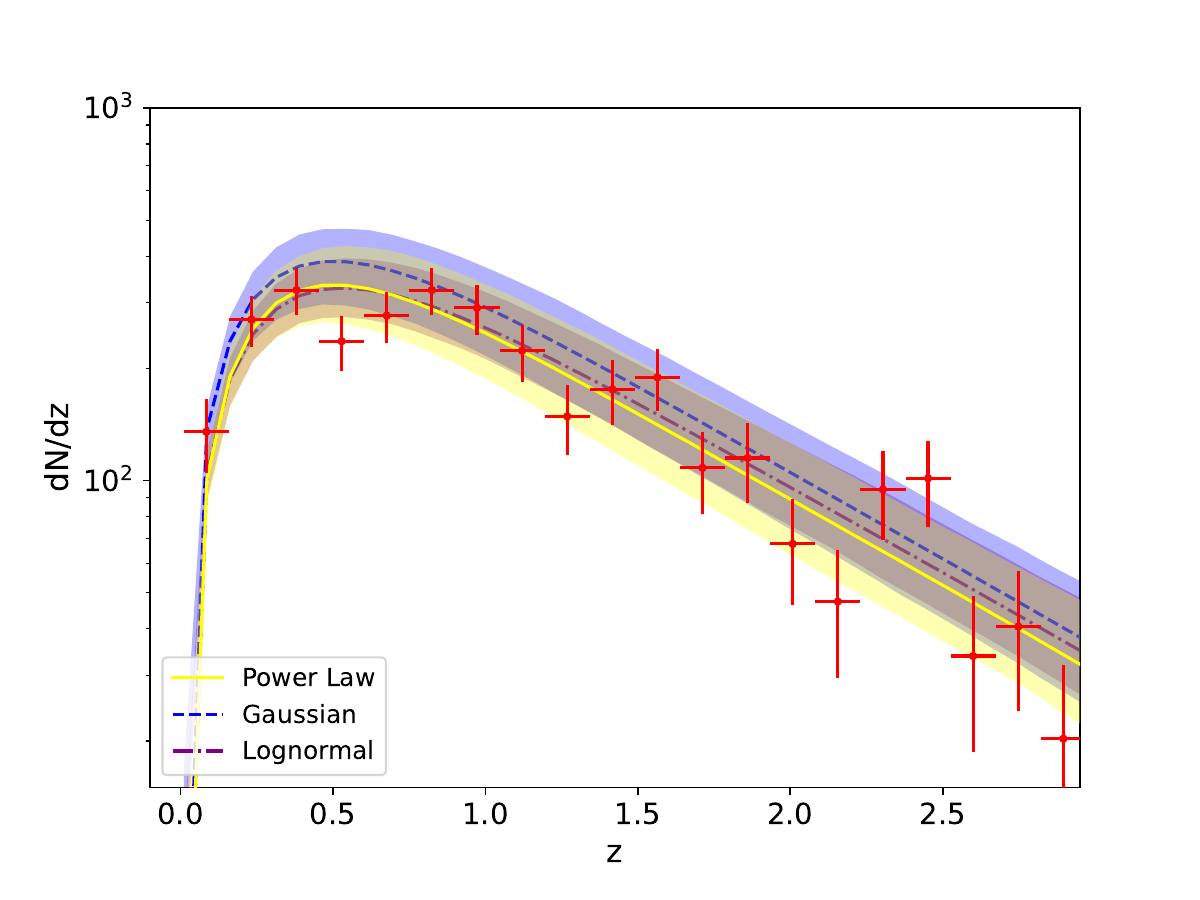} 
\includegraphics[width=0.45\textwidth]{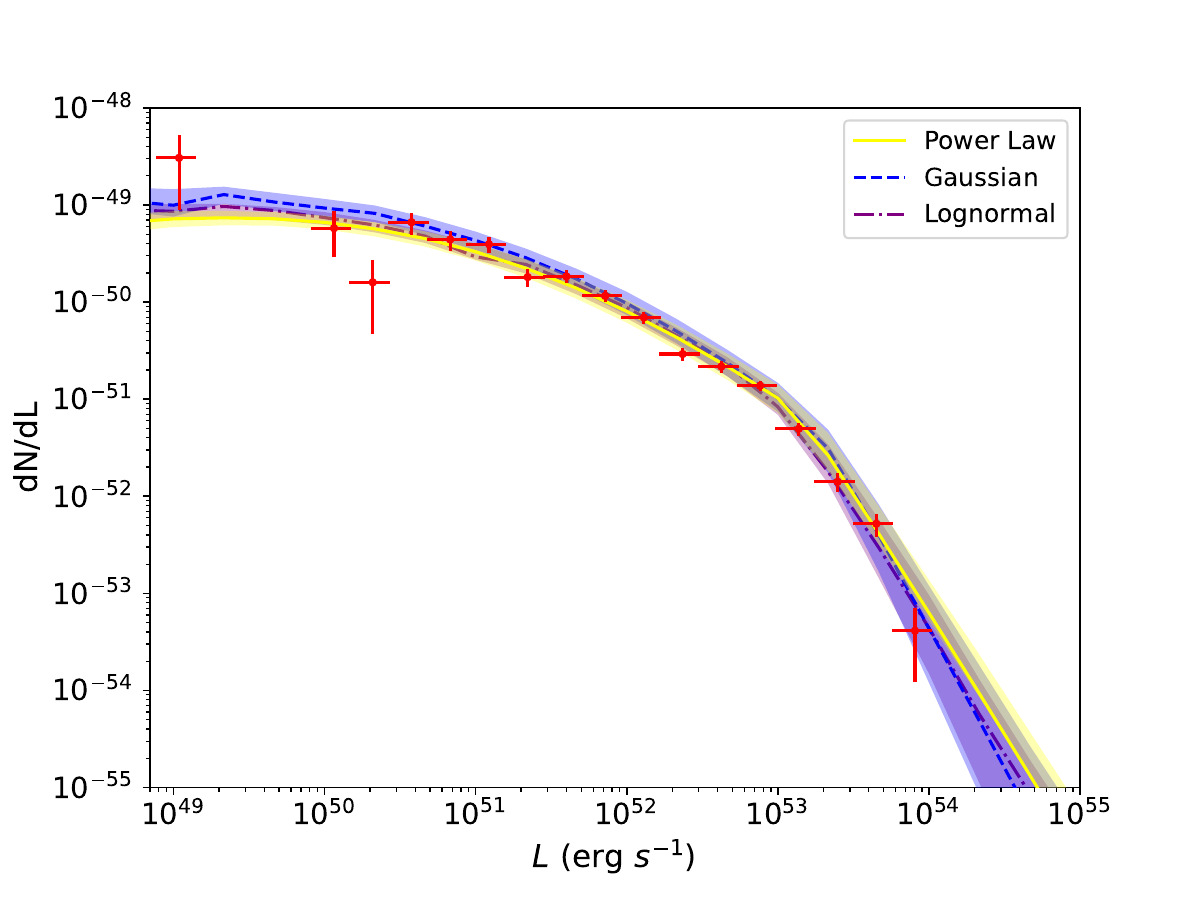}
\caption{Redshift and luminosity distributions of 478 $Fermi$ sGRBs with $P\geq4.1$ $\mathrm{photons\,cm^{-2}\,s^{-1}}$ and
$z\leq3$ (the red points, with the detection number in each redshift or luminosity bin, are indicated by a solid point with 
Poisson error bars). Different curves correspond to the expected distributions from various best-fit delay time distribution models: power-law delay (yellow solid lines), Gaussian delay (blue dashed lines), and lognormal delay (purple dot-dashed lines). Shaded areas show the 1$\sigma$ confidence regions of the corresponding models.}\label{fig:dN}
\end{figure*}

Figure~\ref{fig:dN} presents the redshift and luminosity distributions of 478 sGRBs with $P\geq4.1$ 
$\mathrm{photons\,cm^{-2}\,s^{-1}}$ and $z\leq3$. To facilitate a direct comparison among different 
delay-time distribution models, we show in Figure~\ref{fig:dN} the best-fit theoretical curves 
for the power-law delay model (yellow solid), the Gaussian delay model (blue dashed), and the lognormal delay
model (purple dot-dashed). One can see from this plot that the expectations from these three models appear to 
fit the observed $z$ and $L$ distributions comparably well. According to the DIC model selection criterion,
we find that the Gaussian delay model is statistically preferred with an approximate relative probability of $99.99\%.$

With the best-fit parameters presented in Table~\ref{tab:bestfit}, we are able to calculate the local sGRB
formation rate at $z=0$, $R_{\mathrm{sGRB}}(0)$, using Equation~(\ref{eq:formation}). 
We find
$R_{\mathrm{sGRB}}(0)=1.37_{-0.27}^{+0.30}$ $\mathrm{Gpc^{-3}\,yr^{-1}}$ for the Gaussian delay model,
$1.63_{-0.27}^{+0.44}$ $\mathrm{Gpc^{-3}\,yr^{-1}}$ for the lognormal delay model, and $1.78_{-0.17}^{+0.12}$
$\mathrm{Gpc^{-3}\,yr^{-1}}$ for the power-law delay model, which are consistent with each other within 
the $1\sigma$ confidence level $1\sigma$. We would like to stress that our derived local sGRB formation rate 
is broadly consistent with previous results \citep{gp2005,Yonetoku2014,gir2016,p2018}.
The event rate of BNS mergers is estimated to be as high as $R_{\rm BNS}=320_{-240}^{+490}$ 
$\mathrm{Gpc^{-3}\,yr^{-1}}$, as inferred from the second LIGO-Virgo GW transient catalog \citep{2021ApJ...913L...7A}. 
The discrepancy between the local sGRB formation rate and BNS merger rate arises primarily from the high 
collimation of GRB jets. For top-hat jets, the beaming factor is calculated as 
$f_b=R_{\mathrm{sGRB}}(0)/R_{\rm BNS}$, yielding a half-opening angle of $\theta_j\sim3^{\circ}-12^{\circ}$. 
This range is in good agreement with previous estimates of $\theta_j\sim3^{\circ}-26^{\circ}$ \citep{f2013,f2015,ma2012}.

In our research, the inferred sGRB LF exponents $\nu_1$ and $\nu_2$ range from 1.25 to 1.44 and 2.33 to 3.40, 
respectively, at the 1$\sigma$ confidence level. We emphasize that if the LF is defined in terms of $\phi(L)dL$ 
rather than $\phi(L)d\log L$, our derived LF exponents are roughly consistent with previous studies: 
$\nu_1=1.6\pm0.4$ and $\nu_2=3.0\pm1.0$ by \cite{2006A&A...453..823G}, $\nu_1=1.95_{-0.12}^{+0.12}$ and
$\nu_2=3.0_{-0.8}^{+1.0}$ by \cite{wp2015}, and $\nu_1=1.29\pm0.01$ and $\nu_2=2.07\pm0.01$ by \cite{zhang2018}.

\section{Joint GW and sGRB Detection}
\label{sec:joint}
In this section, we will first describe the synthetic BNS population and provide estimates of the GW detection rates 
by the GW detector networks. Next, we will calculate the sGRB detection rates for high-energy satellites. Finally, 
we will study the joint detection prospects of GWs and their sGRB counterparts from BNS mergers.

\subsection{GW detection rate simulation}
Following \cite{b2024}, we inject $N_{\rm inj}=300,000$ sources from BNS mergers up to a luminosity distance of
$d_{L}=1.6\,\mathrm{Gpc}$ (corresponding to a comoving distance of $\sim1.2\,\mathrm{Gpc}$ and redshift $z\sim0.3$).
The number of injected sources is oversampled by a factor of $\sim130$ compared to the merger rate 
of BNSs within the same comoving volume, based on the median BNS merger event rate of 
$320\,\mathrm{Gpc^{-3}\,yr^{-1}}$ \citep{2021ApJ...913L...7A}. This oversampling is
conducted to avoid any effects of small number statistic on our results. The injected sources are uniformly distributed 
in the rest-frame comoving volume, which translates to a redshift probability distribution of 
\begin{equation}
\frac{\mathrm{d}p(z)}{\mathrm{d}z} \propto \frac{1}{1+z}\frac{\mathrm{d}V(z)}{\mathrm{d}z}\;.
\end{equation}
We assume that the mass distribution of the NSs obeys a normal distribution with a mean of 1.33 $\mathrm{M_{\odot}}$ 
and a standard deviation of 0.09 $\mathrm{M_{\odot}}$ \citep{2016ARA&A..54..401O}. The spins of the NSs are uniformly
distributed between $0 < \chi < 0.05$, with the upper limit set by the maximum known NS spin \citep{2018PhRvD..98d3002Z}.
The direction of the orbital angular momentum is uniformly distributed in space. We generate the GW waveforms using 
the Python package PyCBC\footnote{\url{https://pycbc.org/}} with the IMPRPhenomPv2 approximant model 
\citep{Schmidt12,Hannam14,Khan16}.

\begin{figure*}
    \includegraphics[width=0.3\textwidth]{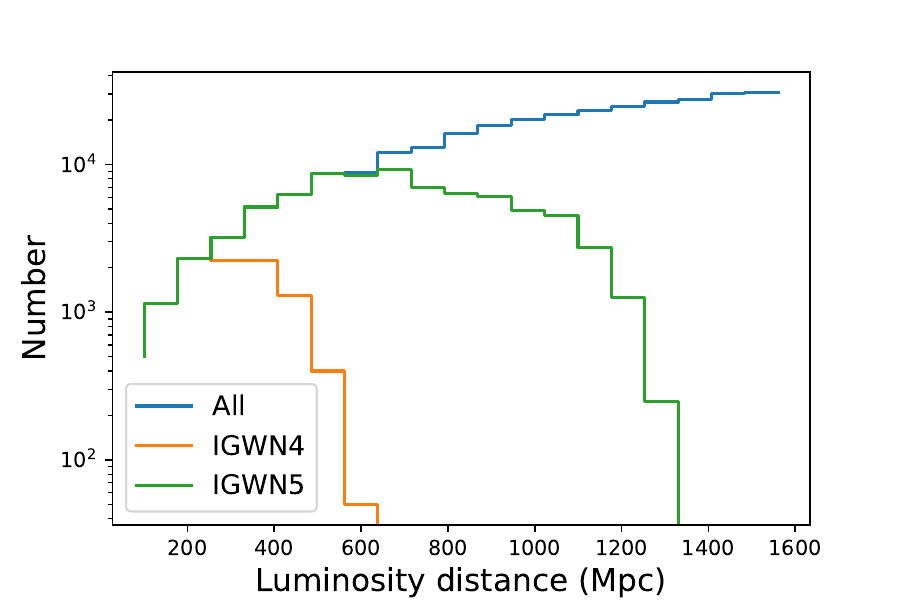}
    \includegraphics[width=0.3\textwidth]{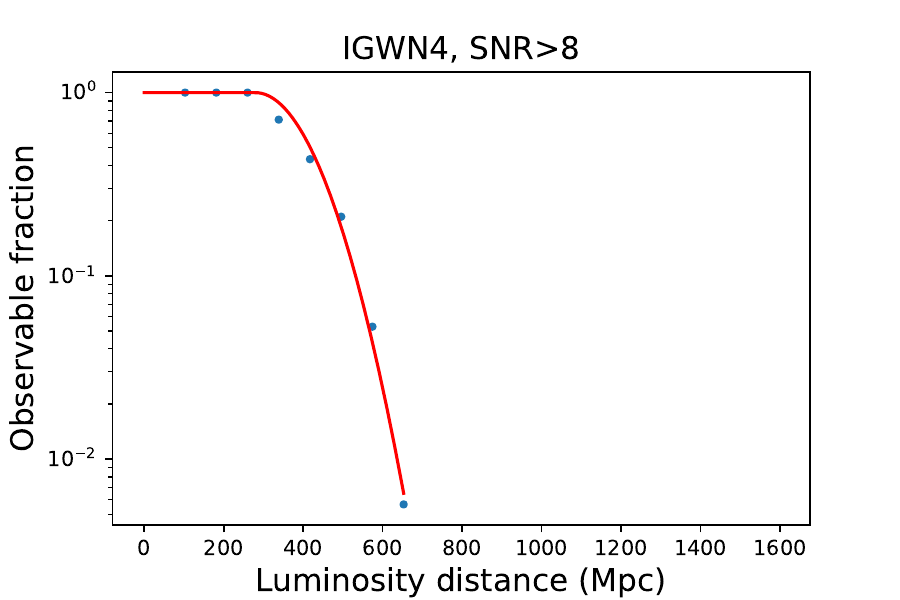}
    \includegraphics[width=0.3\textwidth]{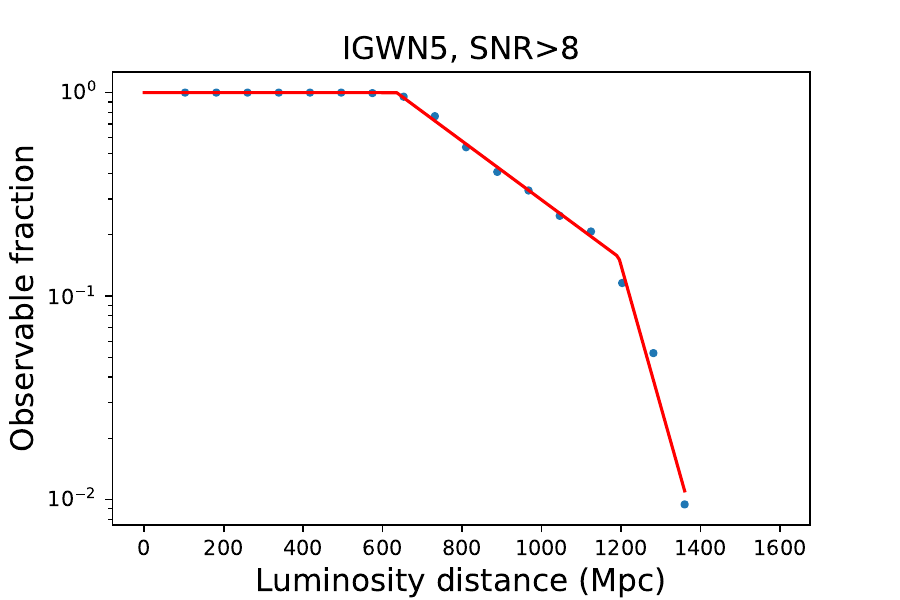}
    \caption{Left panel: Luminosity distance  distribution of injected BNS merger events (blue histogram).
    The yellow and green histograms correspond to the distributions of events detected above the
    threshold $\mathrm{SNR_{net}}$ of 8 by IGWN4 and IGWN5, respectively. Middle panel: GW detection rate
$R_{\mathrm{GW}}(z)$ as a function of luminosity distance, calculated as the ratio of IGWN4 detections to the total injected events at each redshift. The best-fit curve is described by Equation~(\ref{eq:IGWN4}). Right panel: Same as the middle panel but for IGWN5, with the best-fit curve parametrized by Equation~(\ref{eq:IGWN5}).}
    \label{fig:GWdet}
\end{figure*}

To determine which injected BNS merger events can be detected by two different GW detection networks, IGWN4 and IGWN5, 
we consider those events above a particular threshold value of the network signal-to-noise ratio, i.e.,
\begin{equation}
    \mathrm{SNR_{net}}=\sqrt{\sum_{\mathrm{det}}\mathrm{SNR_{det}^{2}}}\,,
\end{equation}
where $\mathrm{SNR_{det}}$ is the signal-to-noise ratio (SNR) of a single detector \citep{2001PhRvD..64d2004P}. 
In this work, we define the criterion of $\mathrm{SNR_{net}}>8$ as the detection threshold \citep{2022ApJ...924...54P}.
For a given single detector, the $\mathrm{SNR_{det}}$ could be calculated by \citep{lisa2021}
\begin{equation}
    \label{eq:snr}
    \mathrm{SNR_{det}^2}=4\operatorname{Re}\left(\int_{f_{\min}}^{f_{\max}}df\frac{\tilde{X}\tilde{X}^*}{S_n(f)}\right)\,,
\end{equation}
where $\tilde{X}$ represents the complex GW signal, and $S_n(f)$ is the noise power spectral density of the corresponding detector, which can be downloaded from the Sensitive Curves website\footnote{\url{https://git.ligo.org/sensitivity-curves/observing-scenario-paper-2019-2020-update/-/tree/master}}. 

From the above simulation, we obtain the distribution of the injected BNS merger events over luminosity distance $d_{l}(z)$, 
which is presented in the left panel of Figure~\ref{fig:GWdet} as the blue histogram. In this plot, we also show 
the distribution of the events detected above the threshold $\mathrm{SNR_{net}}$ of 8 by either IGWN4 (yellow histogram) 
or IGWN5 (green histogram). The middle (right) panel of Figure~\ref{fig:GWdet} displays the GW detection rate
$R_{\mathrm{GW}}(z)$ as a function of redshift (or $d_{l}$), calculated by dividing 
the number of detections from IGWN4 (IGWN5) by the total injected events at each redshift. 
Here, we model $R_{\mathrm{GW}}(z)$ with empirical piecewise functions.
For the case of IGWN4, we have
\begin{equation}
\label{eq:IGWN4}
\begin{aligned}
\log_{10}&R_{\mathrm{GW}} (z)=\\
&\left\{\begin{array}{l}
0;\,\,\,\,\,\,\,\,\,\,\,\,\,\,\,\,\,\,\,\,\,\,\,\,\,\,\,\,\,\,\,\,\,\,\,\,\,\,\,\,\,\,\,\,\,\,\,\,\,\,\,\,\,\,\,\,\,\,\,\,\,\,\,\,\,\,\,\,\,\,\,\,\,\,\,\,\,\, d_{l}(z)\leq d_{b}\;, \\
\kappa\left[d_{l}(z)-d_{b}\right]^{2}+\xi\left[d_{l}(z)-d_{b}\right];\, d_{l}(z)> d_{b}\;,
\end{array}\right.
\end{aligned}
\end{equation}
with best-fit parameters $d_{b}=280\,\mathrm{Mpc}$, $\kappa=-1.56\times 10^{-5}$, 
and $\xi=4.04 \times 10^{-8}$.
For IGWN5, a triple power-law parametrizes $\log_{10}R_{\mathrm{GW}}(d_{l})$:  
\begin{equation}
\label{eq:IGWN5}
\begin{aligned}
&\log_{10}R_{\mathrm{GW}} (z)=\\
&\left\{\begin{array}{l}
0;\,\,\,\,\,\,\,\,\,\,\,\,\,\,\,\,\,\,\,\,\,\,\,\,\,\,\,\,\,\,\,\,\,\,\,\,\,\,\,\,\,\,\,\,\,\,\,\,\,\,\,\,\,\,\,\,\,\,\,\,\,\,\,\,\,\,\,\,\,\,\,\,\,\,\,\,\,\,\,\, d_{l}(z)\leq d_{b_{1}}\;, \\
\kappa\left[d_{l}(z)-d_{b_{1}}\right];\,\,\,\,\,\,\,\,\,\,\,\,\,\,\,\,\,\,\,\,\,\,\,\,\,\,\,\,\,\,\,\,\,\,\,\,\,\,\,\,\,\,\,\, d_{b_{1}}<d_{l}(z)\leq d_{b_{2}}\;,\\
\kappa\left[d_{l}(z)-d_{b_{1}}\right]+\xi\left[d_{l}(z)-d_{b_{2}}\right];\, d_{l}(z)> d_{b_{2}}\;,
\end{array}\right.
\end{aligned}
\end{equation}
with best-fit parameters $d_{b_{1}}=635\,\mathrm{Mpc}$, $d_{b_{2}}=1193\,\mathrm{Mpc}$, 
$\kappa=-1.44\times 10^{-3}$, and $\xi=-6.92 \times 10^{-3}$.

\begin{table*}
\renewcommand{\arraystretch}{1.3}
\caption{The energy bands, flux thresholds, Sky-coverage fractions, and SAA-outage fractions of the high-energy emissions.}
\label{tab2}
\resizebox{0.8\textwidth}{!}{
\begin{tabular}{lcccc}
\hline
Mission  & Energy band  &  Flux threshold  &  Sky-coverage fraction  &  SAA-outage fraction \\
         &  (keV)       &                       &    ($f_{\rm sky}$)      &  ($f_{\rm not\_SAA}$)  \\
\hline
\emph{Fermi}/GBM     &  10--1000  & $4.1\,(\mathrm{photons\,cm^{-2}\,s^{-1}})$ &  0.7  &  0.85 \\
\emph{SVOM}/$\mathrm{GRM^{(a)}}$  &  1--1000    & $3.2\,(\mathrm{photons\,cm^{-2}\,s^{-1}})$ &  0.24  &  0.85  \\
\emph{EP}/$\mathrm{WXT^{(b)}}$    &  0.5--4      & $8.9 \times 10^{-10}\,(\mathrm{erg\,cm^{-2}\,s^{-1}})$           &  0.09  & 0.67  \\
\hline
\end{tabular}
}
\begin{description}
  \item[\emph{Note.}] {(a) \cite{wei2016}, (b) \cite{yuan2015}.}
\end{description}
\end{table*}

\begin{table*}
\renewcommand{\arraystretch}{1.6}
\caption{The sGRB and joint sGRB $+$ GW detection rates for all combinations of high-energy GRB missions, GW networks, and 
delay-time distribution models. Columns 2-4 give the sGRB detection rates for the three GRB missions, independent of GW detection. 
Note that these rates are estimated for events with $d_{L}<1.6\,\mathrm{Gpc}$ (corresponding to $z<0.3$), which is 
the distance limit of this work; they should not be confused with the total sGRB detection rates (see the discussion 
at the end of section~\ref{subsec:sGRB rate}).
The last six columns provide the joint sGRB $+$ GW detection rates. For instance, the last cell of this table states that
for the Gaussian delay model, \emph{EP}/WXT $+$ IGWN5 will detect $0.24_{-0.04}^{+0.03}$ events per year. All the results 
are presented at the $1\sigma$ confidence level.}
\label{tab:rate}
\resizebox{\linewidth}{!}{\begin{tabular}{cccccccccc}\hline
Delay model & \multicolumn{3}{c}{sGRB detection rate $(\mathrm{yr^{-1}})$}                                                                        & \multicolumn{3}{c}{IGWN4 joint detection rate (yr$^{-1}$)}                                                       & \multicolumn{3}{c}{IGWN5 joint detection rate (yr$^{-1}$)}                                                                             \\ \hline
                  & \multicolumn{1}{c}{\emph{Fermi}/GBM}                          & \multicolumn{1}{c}{\emph{SVOM}/GRM}             & \emph{EP}/WXT                    & \multicolumn{1}{c}{\emph{Fermi}/GBM}              & \multicolumn{1}{c}{\emph{SVOM}/GRM}           & \multicolumn{1}{c}{\emph{EP}/WXT}                  & \multicolumn{1}{c}{\emph{Fermi}/GBM}              & \multicolumn{1}{c}{\emph{SVOM}/GRM}           & \multicolumn{1}{c}{\emph{EP}/WXT}                 \\ 
Gaussian          & \multicolumn{1}{c}{$\mathrm{3.28_{-0.58}^{+0.48}}$} & \multicolumn{1}{c}{$2.69_{-0.40}^{+0.32}$}  & $0.78^{+0.09}_{-0.11}$ & \multicolumn{1}{c}{$0.24_{-0.03}^{+0.03}$} & \multicolumn{1}{c}{$0.09_{-0.01}^{+0.01}$} & $0.02_{-0.01}^{+0.01}$ & \multicolumn{1}{c}{$1.18_{-0.21}^{+0.17}$} & \multicolumn{1}{c}{$0.69_{-0.12}^{+0.10}$} & \multicolumn{1}{c}{$0.24_{-0.04}^{+0.03}$} \\ 
Lognormal         & \multicolumn{1}{c}{$2.82_{-0.31}^{+0.22}$}          & \multicolumn{1}{c}{$2.06_{-0.22}^{+0.16}$} & $0.76_{-0.06}^{+0.08}$ & \multicolumn{1}{c}{$0.21_{-0.02}^{+0.03}$} & \multicolumn{1}{c}{$0.08_{-0.01}^{+0.01}$} & $0.02_{-0.01}^{+0.01}$ & \multicolumn{1}{c}{$1.03_{-0.10}^{+0.19}$} & \multicolumn{1}{c}{$0.59_{-0.06}^{+0.10}$} & \multicolumn{1}{c}{$0.21_{-0.02}^{+0.01}$} \\ 
Power law         & \multicolumn{1}{c}{$3.19_{-0.35}^{+0.31}$}           & \multicolumn{1}{c}{$1.99_{-0.25}^{+0.28}$}    & $0.57_{-0.07}^{+0.08}$  & \multicolumn{1}{c}{$0.21_{-0.02}^{+0.02}$} & \multicolumn{1}{c}{$0.08_{-0.01}^{+0.01}$} & $0.02_{-0.01}^{+0.01}$ & \multicolumn{1}{c}{$1.12_{-0.11}^{+0.14}$} & \multicolumn{1}{c}{$0.57_{-0.06}^{+0.06}$} & \multicolumn{1}{c}{$0.16_{-0.01}^{+0.01}$} \\ \hline
\end{tabular}}
\end{table*}

\subsection{sGRB detection rates}
\label{subsec:sGRB rate}
To determine the prospects of sGRB detection, we consider three existing high-energy GRB detectors: \emph{Fermi}/GBM 
(10--1000 keV), the Gamma-Ray Monitor (GRM) of \emph{SVOM} (1--1000 keV), and the Wide-field X-ray Telescope 
(WXT) of \emph{EP} (0.5--4 keV). The energy bands, flux thresholds, sky-coverage fractions, and SAA-outage fractions 
of these three high-energy emissions are summarized in Table~\ref{tab2}.

With the flux threshold and effective detection fraction for each detector, we compute 
the sGRB detection rate per year up to a luminosity distance of $d_{L}=1.6\,\mathrm{Gpc}$ (corresponding to 
$z=0.3$), consistent with the distance limit applied to BNS merger events. The calculation employs 
Equation~(\ref{eq:N}), with model-free parameters fixed to the values in Table~\ref{tab:bestfit}.
For the Gaussian delay model, our results show that the sGRB detection rate is $\mathrm{3.28_{-0.58}^{+0.48}}$ $\mathrm{yr^{-1}}$ 
for \emph{Fermi}/GBM, $2.69_{-0.40}^{+0.32}$ $\mathrm{yr^{-1}}$ for \emph{SVOM}/GRM, and $0.78^{+0.09}_{-0.11}$ 
$\mathrm{yr^{-1}}$ for \emph{EP}/WXT. For the same detector, similar sGRB detection rates are derived for different 
delay-time distribution models. The derived sGRB detection rates for the high-energy GRB detectors are listed 
in Columns 2-4 of Table~\ref{tab:rate}. It is clear that \emph{Fermi}/GBM has the highest sGRB detection rate. 
Due to their relatively low sky coverage, the other two detectors, \emph{SVOM}/GRM and \emph{EP}/WXT, have 
lower detection rates, in that order.

It should be underlined that our analysis is limited by the distance cut-off of $d_{L}=1.6\,\mathrm{Gpc}$
($z=0.3$). Thus, events beyond this distance limit are missed by our calculations. However, in reality, many 
higher-redshift ($z>0.3$) events will be detected by high-energy GRB missions. This makes the sGRB detection 
rates calculated here lower than the true sGRB detection rates by these missions.

Our final \emph{Fermi}/GBM sample contains 478 sGRBs with peak photon flux $P\geq4.1$
$\mathrm{photons\,cm^{-2}\,s^{-1}}$ and redshift $z\leq3$. Over the $15.5\,\mathrm{yr}$ mission period,
the observed \emph{Fermi}/GBM detection rate is $\sim30.8\,\mathrm{yr^{-1}}$. Assuming a \emph{Fermi}/GBM flux 
threshold of $4.1$ $\mathrm{photons\,cm^{-2}\,s^{-1}}$ in the 10--1000 keV energy range, we predict 
a detectable sGRB rate at $z\leq3$ of $34.7_{-5.8}^{+8.0}\,\mathrm{yr^{-1}}$ (Gaussian delay model), 
$30.1_{-3.4}^{+3.8}\,\mathrm{yr^{-1}}$ (Lognormal delay model), and $29.3_{-8.7}^{+9.7}\,\mathrm{yr^{-1}}$ 
(power-law delay model) using Equation~(\ref{eq:N}). These predictions agree well with the observed detection rate 
($\sim30.8\,\mathrm{yr^{-1}}$) within uncertainties.

\subsection{Joint sGRB and GW detection rates}
By combining the GW detection rate $R_{\mathrm{GW}}(z)$ with the observed sGRB distributions, the expected number of 
joint sGRB and GW detections from BNS mergers can be estimated using specific GRB and GW facilities through 
the following equation:
\begin{equation}
\begin{aligned}
N_{\rm joint}=T f_{\rm sky} f_{\rm not\_SAA}\int_{0}^{z_{\rm lim}}\frac{R_{\rm sGRB}(z)R_{\mathrm{GW}}(z)}{1+z}\frac{\mathrm{d}V(z)}{\mathrm{d}z}\mathrm{d}z\\
\int_{\max[L_{\rm min},\,L_{\rm lim}(z)]}^{L_{\rm max}}\phi(L)\mathrm{d}L \,,
\end{aligned}
\label{eq:N_joint}
\end{equation}
where $z_{\rm lim}=0.3$ (corresponding to $1.6\,\mathrm{Gpc}$) is the distance limit of this work. 
It is important to note that in the calculations above (specifically Equations \ref{eq:N} 
and \ref{eq:N_joint}), all sGRB observations are assumed to be on-axis, with jets modeled under 
a top-hat structure. However, observations of GW170817 and GRB 170817A suggest that GRB jets likely 
follow a wide angular distribution \citep{ab2017}. Because the luminosity of a given sGRB decreases 
rapidly as the viewing angle $\theta$ relative to its symmetry axis increases, off-axis detection 
is feasible only for nearby sources. Consequently, the top-hat assumption remains a reasonable 
approximation for sGRBs at cosmological distances. For nearby events like GRB 170817A, however, 
the jet structure and the potential for off-axis detection become significant.

Because of the on-axis assumption, the sGRB LF $\phi(L)$ derived in Section~\ref{sec:sGRB} describes 
only the distribution of the core luminosity $L$ of GRB jets. For a structured jet, the observed luminosity 
is expressed as
\begin{equation}
    \label{eq:jet}
    L_{\mathrm{obs}}(\theta)=LJ(\theta)\;,
\end{equation}
where $J(\theta)$ is a function describing the viewing-angle $\theta$ dependence of the luminosity, 
and $L$ is the on-axis luminosity at $\theta=0$. The observed LF must instead be expressed as
\begin{equation}
    \label{eq:fdis}
    \phi'(L_{\rm obs}) \propto \int_{0}^{\pi/2}\phi\left[L_{\rm obs}/J(\theta)\right]\mathrm{sin}\theta\mathrm{d}\theta \,.
\end{equation}
To account for structured jets, the expected number of joint sGRB and GW detections from BNS mergers 
is calculated by
\begin{equation}
    \label{eq:nearNjoint}
    \begin{aligned}
N_{\rm joint}=Tf_{\rm sky} f_{\rm not\_SAA}\int_{0}^{z_{\rm lim}}\frac{R_{\rm sGRB}(z)R_{\mathrm{GW}}(z)}{1+z}\frac{\mathrm{d}V(z)}{\mathrm{d}z}\mathrm{d}z\\
\int_{\max[L_{\rm min},\,L_{\rm lim}(z)]}^{L_{\rm max}}\phi'(L_{\rm obs})\mathrm{d}L_{\rm obs} \,.
\end{aligned}
\end{equation}
Following \cite{liu2019}, we adopt a Gaussian jet model $J(\theta)=e^{-(\frac{\theta}{2\theta_c})^2}$, 
where $\theta_c=3^{\circ}$ defines the angular core width. Again, in our calculations, the model-free 
parameters are fixed at the best-fit values given in Table~\ref{tab:bestfit}. 
The joint detection rate is calculated using Equation~(\ref{eq:nearNjoint}) with $T=1\,\mathrm{yr}$.

For the sensitivity of the IGWN4, the overall joint sGRB and GW detection rates are rather low. 
The joint detection rates for the high-energy GRB detectors \emph{Fermi}/GBM, \emph{SVOM}/GRM, 
and \emph{EP}/WXT lie in the range of 0.19--0.27,
0.07--0.10, and 0.01--0.03 $\mathrm{yr^{-1}}$, respectively, for different delay-time distribution models 
(see Columns 5-7 of Table~\ref{tab:rate}). The joint detection rate expected from \emph{Fermi}/GBM and IGWN4 means that
only 1--3 sGRB and GW associated events would be detected in about 10 years of observation, which is consistent with 
the current result that only one joint detection of GW170817/GRB 170817A has been found.
After the GW detection network is upgraded to IGWN5, the joint detection rates increase to 0.93--1.35,
0.51--0.79, and 0.15--0.27 $\mathrm{yr^{-1}}$ for \emph{Fermi}/GBM, \emph{SVOM}/GRM, and \emph{EP}/WXT, respectively
(see Columns 8-10 of Table~\ref{tab:rate}).

Our joint sGRB and GW detection rates align well with those estimated in previous studies.
For instance, \cite{2022ApJ...937...79C} reported a joint detection rate of $\sim0.17\,\mathrm{yr^{-1}}$ 
for \emph{Fermi}/GBM and the LIGO-Virgo-KAGRA network during O4, and \cite{2023A&A...680A..45S} predicted 
0.2--1.3 detectable GW-sGRB events per year for the same instruments. Similarly, \cite{b2024} found rates 
of 0.07--0.62 $\mathrm{yr^{-1}}$ and 0.8--4.0 $\mathrm{yr^{-1}}$ for \emph{Fermi}/GBM paired with IGWN4 
and IGWN5, respectively. Our IGWN4 rates are also broadly consistent with \cite{2019MNRAS.485.1435H} after 
scaling for the higher BNS merger rate assumed in their analysis. However, the study by \cite{p2022} yielded 
a higher rate, which likely stems from either brighter emission models or the inclusion of Gaussian noise 
in GW detector simulations.

\section{Conclusions and discussion}
\label{sec:con}
In this paper, we have updated and expanded the sGRB sample detected by the \emph{Fermi} satellite, which now includes 
553 sGRBs as of December 2023. Due to instrumental limitations in sampling faint bursts, 
we focused on sGRBs that are sufficiently bright in the 10--1000 keV \emph{Fermi}/GBM energy band. Specifically, 
we selected bursts with a 64-ms peak photon flux of $P \geq 4.1$ $\mathrm{photons\,cm^{-2}\,s^{-1}}$. Above this 
flux threshold, the incomplete flux sampling due to instrumental selection effects becomes negligible. Consequently, 
our subsample consists of 512 sGRBs with $P \geq 4.1$ $\mathrm{photons\,cm^{-2}\,s^{-1}}$.
Given the limited number of sGRBs with known redshifts, we estimated the pseudo redshifts of 512 sGRBs using
the empirical $E_{p}$-$L$ correlation. Since the maximum redshift observed for sGRBs is 2.609, 
we excluded those bursts with pseudo-redshifts $z>3$. As a result, we analyzed 478 sGRBs with $P\geq4.1$ 
$\mathrm{photons\,cm^{-2}\,s^{-1}}$ and $z\leq3$.

Using the sample of 478 sGRBs with $P\geq4.1$ $\mathrm{photons\,cm^{-2}\,s^{-1}}$ and $z\leq3$, we investigated 
the sGRB LF and formation rate under various delay-time distribution models. We applied the maximum likelihood method 
to analyze the observed redshift and luminosity distributions of the sGRB sample, and used the MCMC technology to 
determine the best-fit parameters for each model. We found that all delay-time distribution models seem to
account for the observed redshift and luminosity distributions similarly well. Nevertheless, based on the DIC model 
selection criterion, the Gaussian delay model is statistically preferred, with an estimated relative probability of 
$99.9\%$. Additionally, with the best-fit parameters presented in Table~\ref{tab:bestfit}, 
consistent local formation rates of sGRBs are obtained for different delay-time distribution models, i.e., 
$R_{\mathrm{sGRB}}(0)=1.37_{-0.27}^{+0.30}$ $\mathrm{Gpc^{-3}\,yr^{-1}}$ for the Gaussian delay model, 
$1.63_{-0.27}^{+0.44}$ $\mathrm{Gpc^{-3}\,yr^{-1}}$ for the lognormal delay model, and $1.78_{-0.17}^{+0.12}$
$\mathrm{Gpc^{-3}\,yr^{-1}}$ for the power-law delay model. These results are well-aligned with those of other works
\citep{gp2005,Yonetoku2014,gir2016,p2018}.

Furthermore, we explored the prospects for joint detection of sGRBs and their GW counterparts from BNS mergers using 
three high-energy GRB detectors (\emph{Fermi}/GBM, \emph{SVOM}/GRM, and \emph{EP}/WXT) and two GW detector networks
(IGWN4 and IGWN5). In our analysis, the GW detection rates by the GW detector networks were estimated using
the synthetic BNS population \citep{b2024}. By combining the GW detection rate with the observed 
sGRB distributions, we further calculated the joint sGRB and GW detection rates across various GRB and GW facilities,
incorporating the angular dependence of the emission energy in GRB jets.
For the sensitivity of the IGWN4, the overall joint sGRB and GW detection rates are relatively low. 
Our predicted joint detection rates for the GRB detectors \emph{Fermi}/GBM, \emph{SVOM}/GRM, and \emph{EP}/WXT 
with IGWN4 are 0.19--0.27, 0.07--0.10, and 0.01--0.03 $\mathrm{yr^{-1}}$, respectively, for different delay-time 
distribution models (see Columns 5-7 of Table~\ref{tab:rate}). Our results offer a plausible explanation for 
the lack of joint sGRB and GW detection over a long period of observation due to the low joint detection rate of 
current instruments. Following the upgrade of the GW detection network to IGWN5, we expect to detect more sGRB-GW 
association events in the future. Specificlly, with IGWN5, the joint detection rates would increase to 0.93--1.35,
0.51--0.79, and 0.15--0.27 $\mathrm{yr^{-1}}$ for \emph{Fermi}/GBM, \emph{SVOM}/GRM, and \emph{EP}/WXT, respectively
(see Columns 8-10 of Table~\ref{tab:rate}).

In this research, besides \emph{Fermi} and \emph{SVOM}, we also studied the sGRB detection rate of \emph{EP}. 
However, \emph{EP} is specifically designed for X-ray observation, and GRB prompt emission is not the primary aim 
of the satellite. As expected, \emph{EP} does not show a strong capability in the detection of sGRBs. The observed 
luminosity of GRB prompt emission depends on the jet structure. In all the jet structure models, the observed 
luminosity decreases rapidly as the angle between the line of sight and the jet increases. Most GRBs could not be 
detected due to off-axis observation. Differently, the afterglow of GRBs allows for observation from larger angles, 
and it is easier to detect in the X-ray band. However, there are insufficient sGRB afterglow samples to perform 
a study predicting the afterglow detection rate of an X-ray space telescope.

Finally, we note a caveat regarding our methodology. While we adopt the $E_p$-$L$ correlation 
for pseudo-redshift estimation throughout this work, significant systematic uncertainties affect this approach. 
Future refinement of redshift-estimation techniques or larger samples of sGRBs with spectroscopic redshifts 
will enable robust validation of our results.

\section*{Acknowledgements}
We thank the anonymous referee for their constructive comments, 
which greatly improved the clarity of this manuscript.
We are also grateful to Dr. Shi-Jie Zheng, Dr. Guang-Xuan Lan, An Li, and Yi-Fang Liang for useful discussions.
This work is partially supported by the Strategic Priority Research Program of the Chinese Academy
of Sciences (grant No. XDB0550400), the National Key R\&D Program of China (2024YFA1611704),
the National Natural Science Foundation of China (grant Nos. 12422307, 12373053,
and 12321003), and the Natural Science Foundation of Jiangsu Province (grant No. BK20221562).

\section*{Data Availability}
All the GRB data used in this paper can be obtained from the Fermi GBM catalog: \url{https://heasarc.gsfc.nasa.gov/W3Browse/fermi/fermigbrst.html}.



\bibliographystyle{mnras}
\bibliography{example} 




\appendix
\section{Corner Plots of the MCMC inference}

\begin{figure}
	\includegraphics[width=0.48\textwidth]{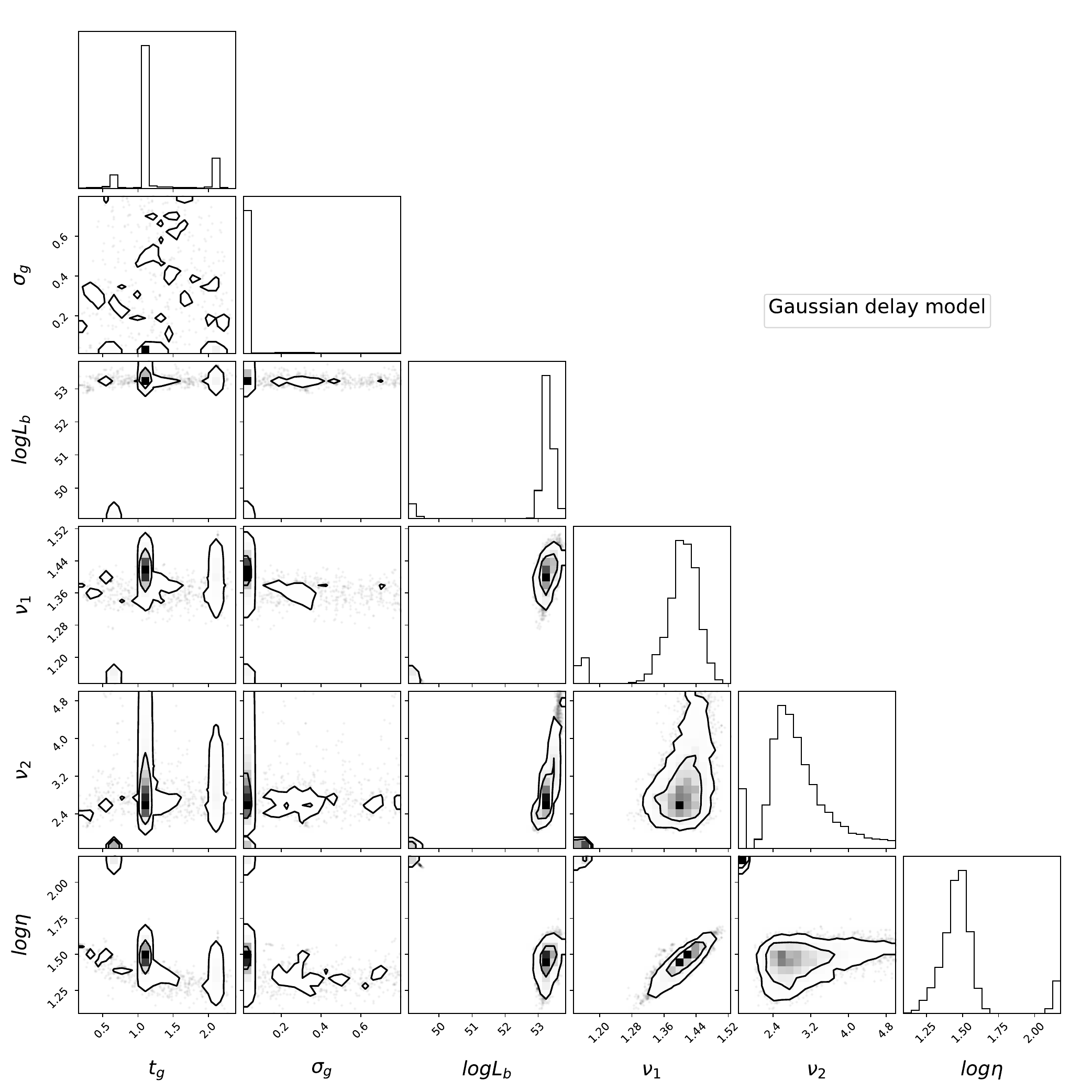}
    \caption{
    1D posterior distributions and 2D confidence regions 
    (with 1-2$\sigma$ contours) for the parameters of the Gaussian delay model.}
    \label{fig:cornerGauss}
\end{figure}

\begin{figure}
    \includegraphics[width=0.48\textwidth]{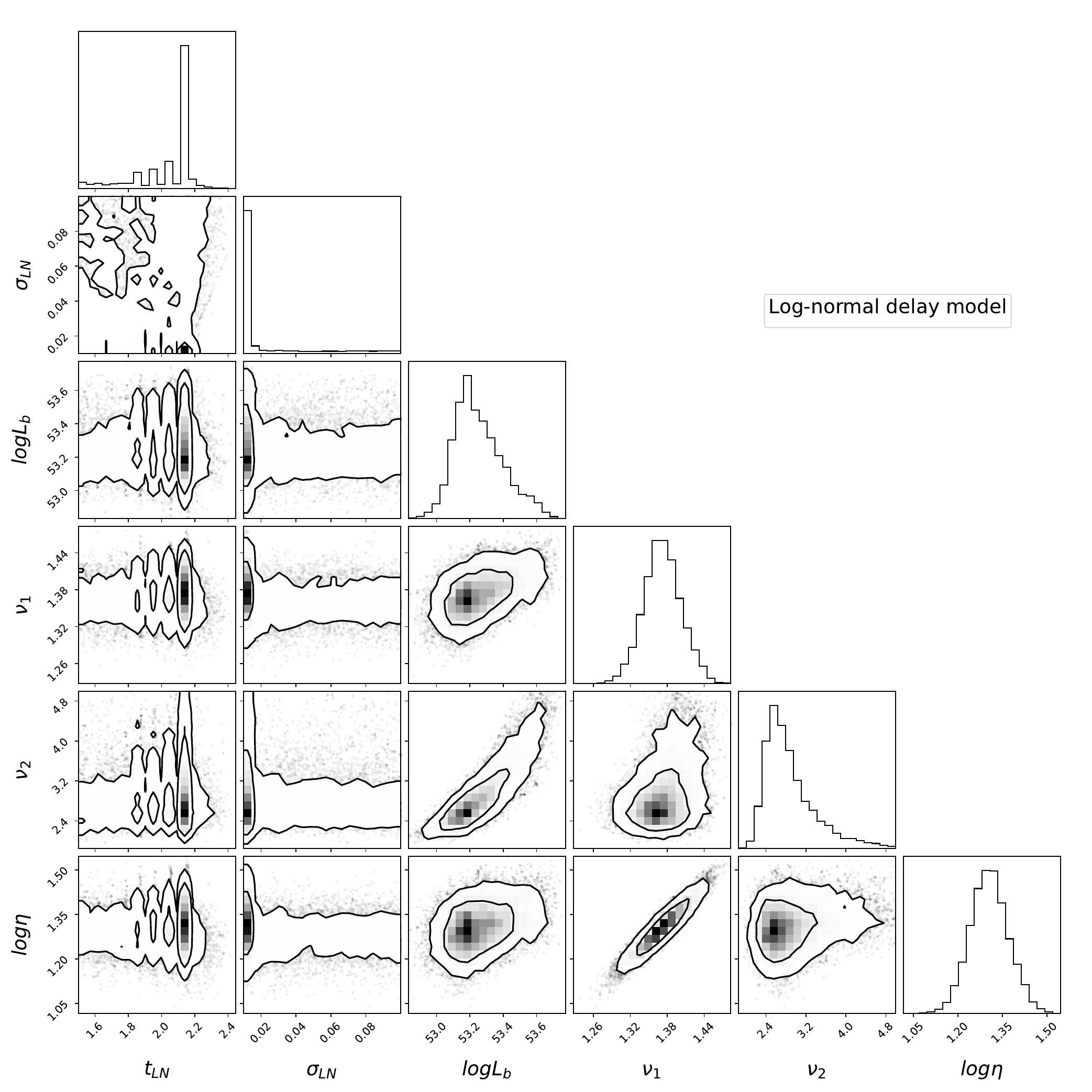}
    \caption{Same as Figure~\ref{fig:cornerGauss}, 
    but for the parameters of the Lognormal delay model.}
    \label{fig:cornerLN}
\end{figure}

\begin{figure}
    \includegraphics[width=0.45\textwidth]{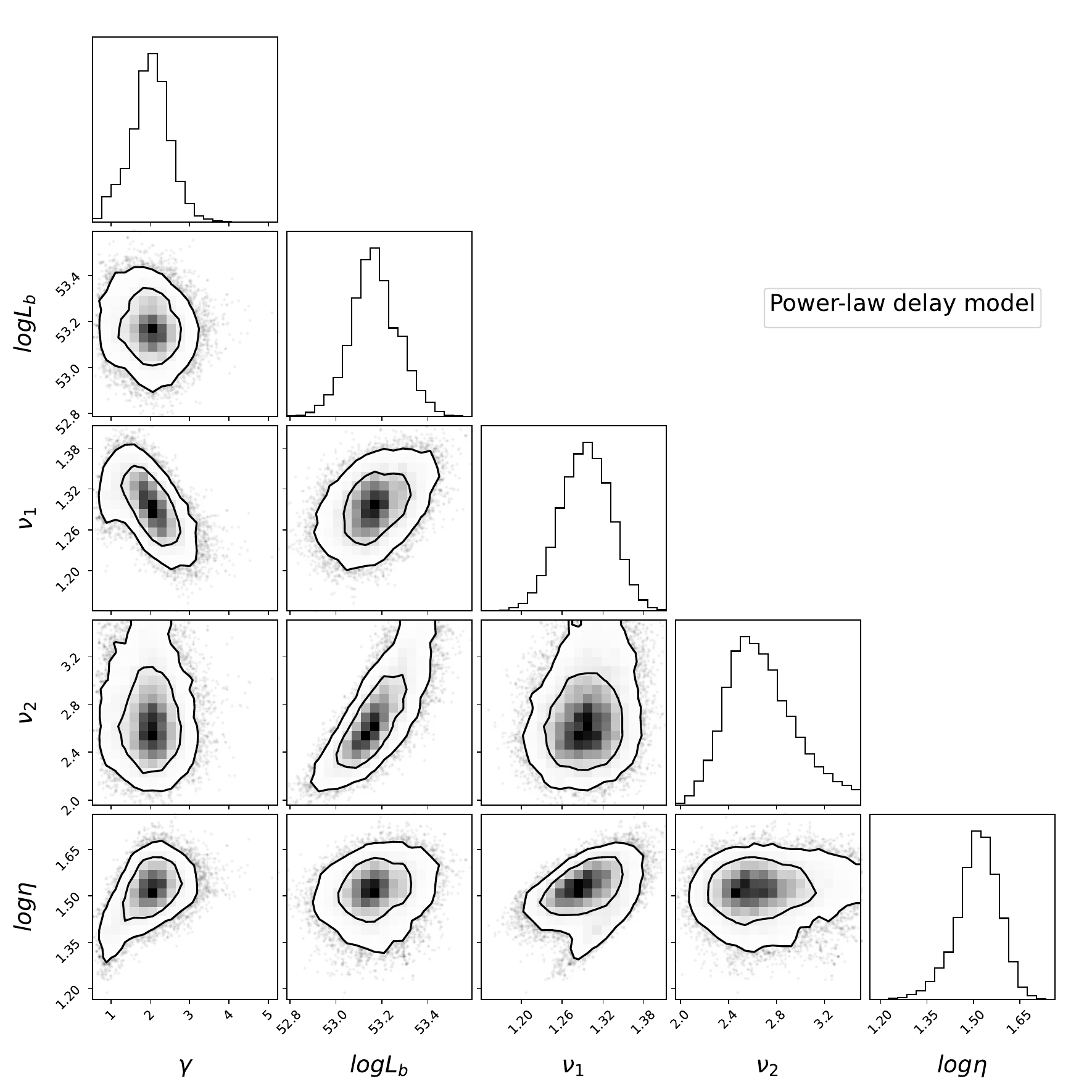}
    \caption{Same as Figure~\ref{fig:cornerGauss}, 
    but for the parameters of the power-law delay model.}
    \label{fig:cornerPL}
\end{figure}




\bsp	
\label{lastpage}
\end{document}